\documentclass[twocolumn,prl,showpacs,amsfonts,amsmath,assymb,eufrak]{revtex4}

\usepackage{epsfig}
\usepackage{color}
\usepackage{bm}

\usepackage{braket}

\usepackage{graphicx}

\newcommand{\eq}[1]{\begin{equation} #1 \end{equation}}
\newcommand{\eqa}[2]{\begin{equation} #1 \label{#2} \end{equation}}
\newcommand{\balign}[1]{\begin{align} #1 \end{align}}
\newcommand{\bcases}[1]{\begin{cases} #1 \end{cases}}

\newcommand{\figin}[4]
{\begin{figure}[tb]
\centering
\includegraphics[width= #1]{#2.pdf}
\caption{#3}
\label{f:#4}
\end{figure}}

\newcommand{\todayd}{\the\year/\the\month/\the\day}

\newcommand{\bib}{\bibitem}

\newcommand{\lmd}{\lambda}

\newcommand{\lb}{\label}
\newcommand{\nt}{\notag}

\newcommand{\bel}{\begin{easylist}}
\newcommand{\eel}{\end{easylist}}

\newcommand{\eref}[1]{Eq.~\eqref{#1}}

\newcommand{\fref}[1]{Fig.~\ref{f:#1}}

\def \({\left(}
\def \){\right)}

\newcommand{\abs}[1]{\left|#1\right|}

\newcommand{\sumtwo}[2]%
{\mathop{\sum_{#1}}_{#2}}
\newcommand{\sumthree}[3]%
{\mathop{\mathop{\sum_{#1}}_{#2}}_{#3}}
\newcommand{\sumfour}[4]%
{\mathop{\mathop{\mathop{\sum_{#1}}_{#2}}_{#3}}_{#4}} 
\newcommand{\prodtwo}[2]%
{\mathop{\prod_{#1}}_{#2}}
\newcommand{\mintwo}[2]%
{\mathop{\min_{#1}}_{#2}}
\newcommand{\maxtwo}[2]%
{\mathop{\max_{#1}}_{#2}}
\newcommand{\maxthree}[3]%
{\mathop{\mathop{\max_{#1}}_{#2}}_{#3}}
\newcommand{\limtwo}[2]%
{\mathop{\lim_{#1}}_{#2}}
\newcommand{\suptwo}[2]%
{\mathop{\sup_{#1}}_{#2}}
\newcommand{\supthree}[3]%
{\mathop{\mathop{\sup_{#1}}_{#2}}_{#3}}
\newcommand{\supfour}[4]%
{\mathop{\mathop{\mathop{\sup_{#1}}_{#2}}_{#3}}_{#4}} 
\newcommand{\inftwo}[2]%
{\mathop{\inf_{#1}}_{#2}}
\newcommand{\infthree}[3]%
{\mathop{\mathop{\inf_{#1}}_{#2}}_{#3}}
\newcommand{\inffour}[4]%
{\mathop{\mathop{\mathop{\inf_{#1}}_{#2}}_{#3}}_{#4}} 

\newcommand{\ep}{\varepsilon}

\newcommand{\para}[1]{{\em #1}\/.---}
\newcommand{\Ncc}{N_{\rm cc}}
\def\no{n^{\rm o}}
\def\ne{n^{\rm e}}
\def\nv{n^{\rm v}}

\def\rnum#1{\resizebox{0.5em}{\height}{\expandafter{\romannumeral #1}}}
\def\Rnum#1{\resizebox{0.5em}{\height}{\uppercase\expandafter{\romannumeral #1}}}

\newcommand{\titlename}{Constructing Concrete Hard Instances of the Maximum Independent Set Problem}

\begin{document}

\preprint{APS/123-QED}

\title{\titlename}

\author{Naoto Shiraishi}
\affiliation{Department of Physics, Keio university, Hiyoshi 3-14-1, Kohoku-ku, Yokohama, Japan}%

\author{Jun Takahashi}
\affiliation{Department of Basic Science, University of Tokyo, Komaba 3-8-1, Meguro-ku, Tokyo, Japan}%
\altaffiliation[Current affiliation:]{~Institute of Physics, Chinese Academy of Sciences, No.8, 3rd South Street, Zhongguancun, Beijing, China}
\email[]{jt@iphy.ac.cn}

\date{\today}

\begin{abstract}
We provide a deterministic construction of hard instances for the maximum independent set problem (MIS).
The constructed hard instances form an infinite graph sequence with increasing size, which possesses similar characteristics to sparse random graphs and in which MIS cannot be solved efficiently.
We analytically and numerically show that all algorithms employing cycle-chain refutation, 
which is a general refutation method we introduce for capturing the ability of many known algorithms, 
cannot upper bound the size of the maximum independent set tightly. 

\end{abstract}

\pacs{
}

\maketitle

\para{Introduction}
Hardness of optimization problems has been an important topic not only in computer science but also in physics, cryptography theory, and engineering.
Traditional computational complexity theory in computer science mainly concerns the {\it worst-case hardness}, 
where many significant results including NP-completeness and hardness of approximation have been developed~\cite{ABbook}.
Recently, the cooperation of computer science and statistical physics has shed light on the problems of the {\it average-case hardness}, in which we consider the hardness of random instances of the problem.
From the side of statistical mechanics, similarity between some optimization problems and some physical models were first pointed out~\cite{MP85, MP86, FA}, 
and then the easy-to-hard transition threshold where random instances become typically easy/hard has been evaluated~\cite{CKT, MSL, KS, MZ97, Mon99,  LWZ, WH, MPZ, MZ, HWbook} with some mathematical supports~\cite{Fri99, AP, ANP, BGT, DSS}.
In addition, similarity to the spin glass transition phenomena including replica symmetry breaking and the complex structure of the state space has also been studied intensively~\cite{Zhou05, Krz07, ZK07, AC, Zde, Li09, Zhou, MMbook, ZK10, Bra}.

In spite of these successes, one of the most important questions, {\it why a hard problem is hard}, has not yet been fully addressed.
Compared to past studies with the physical approaches, the computational complexity approach has not been successful on understanding the average case hardness ~\cite{BT}.
Contrary to the situation for worst case hardness where numerous hard problems could be reduced to each other forming a huge class of NP-complete, reductions regarding average case hard problems is found only in very few cases ~\cite{Lip, Ajt, Reg, DLS, WBP}.
One stumbling block to clarifying the origin of hardness is that we do not have concrete examples of hard instances. 
While there have been attempts to obtain hard instances mainly for a benchmark of some new algorithms~\cite{Bar, Xu, JMS, Kat, Hen, MMH, THH}, these constructions are probabilistic, not deterministic, and the hardness of instances is usually confirmed only empirically.
In a probabilistic construction, we know that most of the instances are hard, but we cannot specify a concrete instance as a hard instance. 
Therefore, constructing a concrete hard instance which reflects properties of typical hard instances will be of great importance for a further understanding of the origin of hardness in hard problems.

In this Letter, we construct the first example of concrete hard instances of an optimization problem called the {\it maximum  independent set} problem (MIS).
We provide a construction of an infinite sequence of graphs with increasing size, which is deterministic, not probabilistic.
An algorithm exactly solving an optimization problem carries out two tasks; searching and refuting, and we focus on the hardness of the latter refutation process.
The average hardness of refutation for another optimization problem was previously conjectured by Feige, with which some optimization problems are shown to be hard for approximation in the sense of worst-case hardness~\cite{Fei, Ale}.
By fixing the employed algorithm to a specific type, e.g., Sum-of-Squares method and semidefinite-programing method, the possibility and impossibility of average hardness of refutation was proven for some constraint satisfaction problems~\cite{BM, ADW, KMDW, DMDSS}.
Following this line, we fix the method of refutation to a specific type which covers many of known algorithms for MIS, and show that refutation is impossible for the constructed instances.
We first derive a lower bound on possible refutation for the constructed instances, and then show numerically and analytically that the true optimal value for these instances is strictly less than that.

\figin{8.5cm}{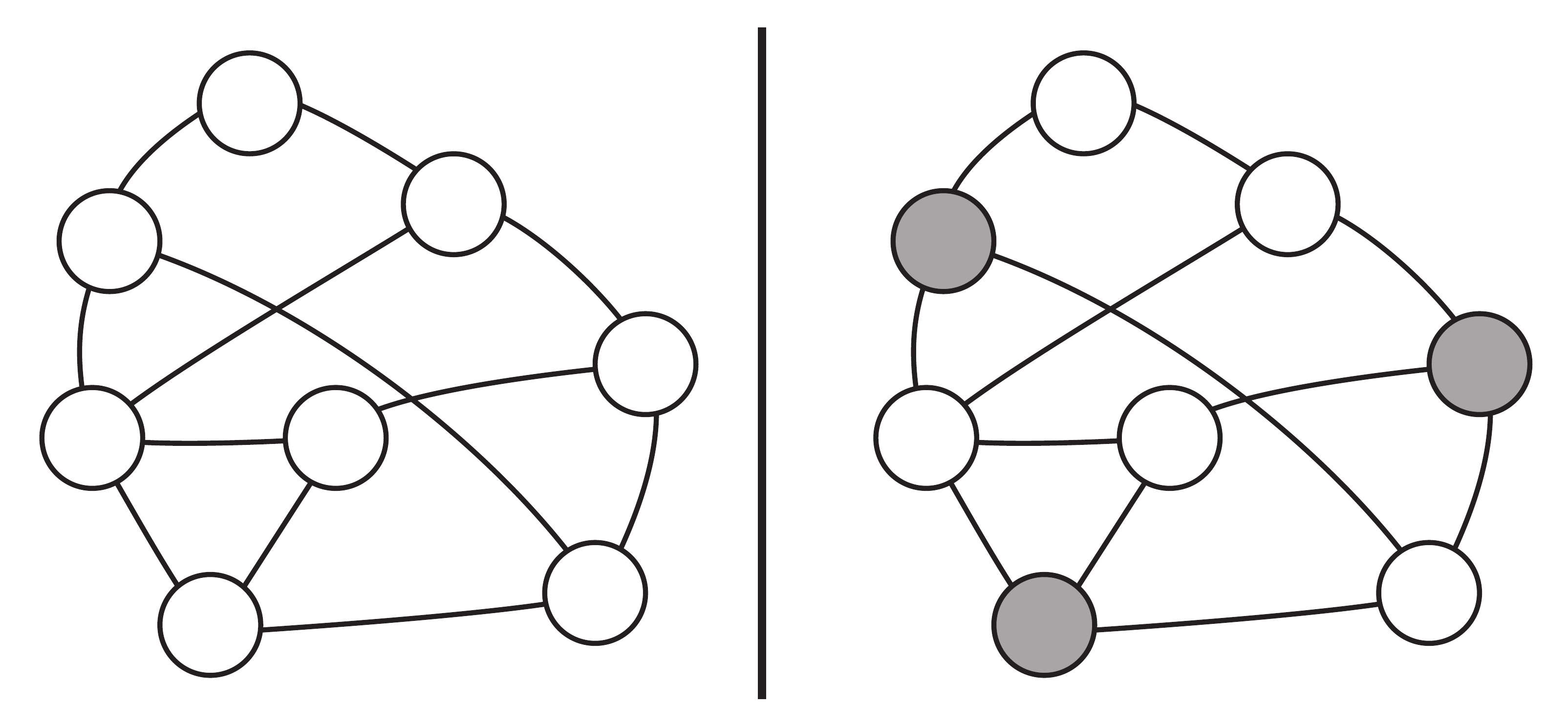}{
Left: An example of a graph.
In MIS, we seek the largest set of vertices which are not neighboring with each other.
Right: One of the maximum independent sets drawn in gray (three vertices) of this graph.
}{mis-ex}

\para{Maximum independent set problem}
We first explain the maximum independent set problem (MIS).
For a given graph $G$, an {\it independent set} is a set of vertices which are not neighboring with each other on the graph (i.e., for any edge at least one of two end point vertices is not included in the set).
The task of MIS is to find (one of) the largest independent set(s).
For example, given the graph in the left of \fref{mis-ex}, the set of gray vertices in the right figure is one of the largest independent sets.
The size of the largest independent set is called the {\it independent number}, which we denote by $N^*$.
We define {\it independent ratio} as $N^*/N$, where $N$ denotes the number of all vertices in the given graph.

It is known that MIS is an NP-complete problem \cite{MIS-NPness}.
Numerical simulations imply that MIS for Erd\"{o}s-Renyi random graphs with the average degree larger than Napier's constant $2.718\cdots$ is hard to compute efficiently on average~\cite{WH, TH14, ore}.
In other words, even the best known algorithms fail to solve MIS of a random graph with high probability, if computation time is bounded by a polynomial of the graph size $N$.
However, perhaps surprisingly, we had no concrete instance of a graph which is indeed hard to solve MIS, although most random graphs are hard.
One difficulty for finding {\it hard instances} is that one could always construct an algorithm which solves a particular instance in short time by simply hard-wiring this instance~\cite{hard-wire}, if the number of those hard instances is finite.
Thus, the aim of this Letter is to provide explicit and deterministic construction of an infinite graph sequence, which contains provably hard instances with arbitrarily large size.

An exact algorithm for MIS should handle the following two tasks.
(i) {\it Searching}: to find an independent set with size $N^*$.
(ii) {\it Refuting}: to ensure that the size of all other independent sets are less than or equal to $N^*$.
When we consider the large size limit, we can generalize our result to allow $o(N)$ errors, that is, the algorithms is regarded as succeeded if it finds an independent set with size $N^*-o(N)$ and it refutes an independent set with size $N^*+o(N)$.
We focus on the hardness of the latter task, refutation, and construct a graph sequence such that all plausible algorithms explained below fail to refute. 

\figin{8.5cm}{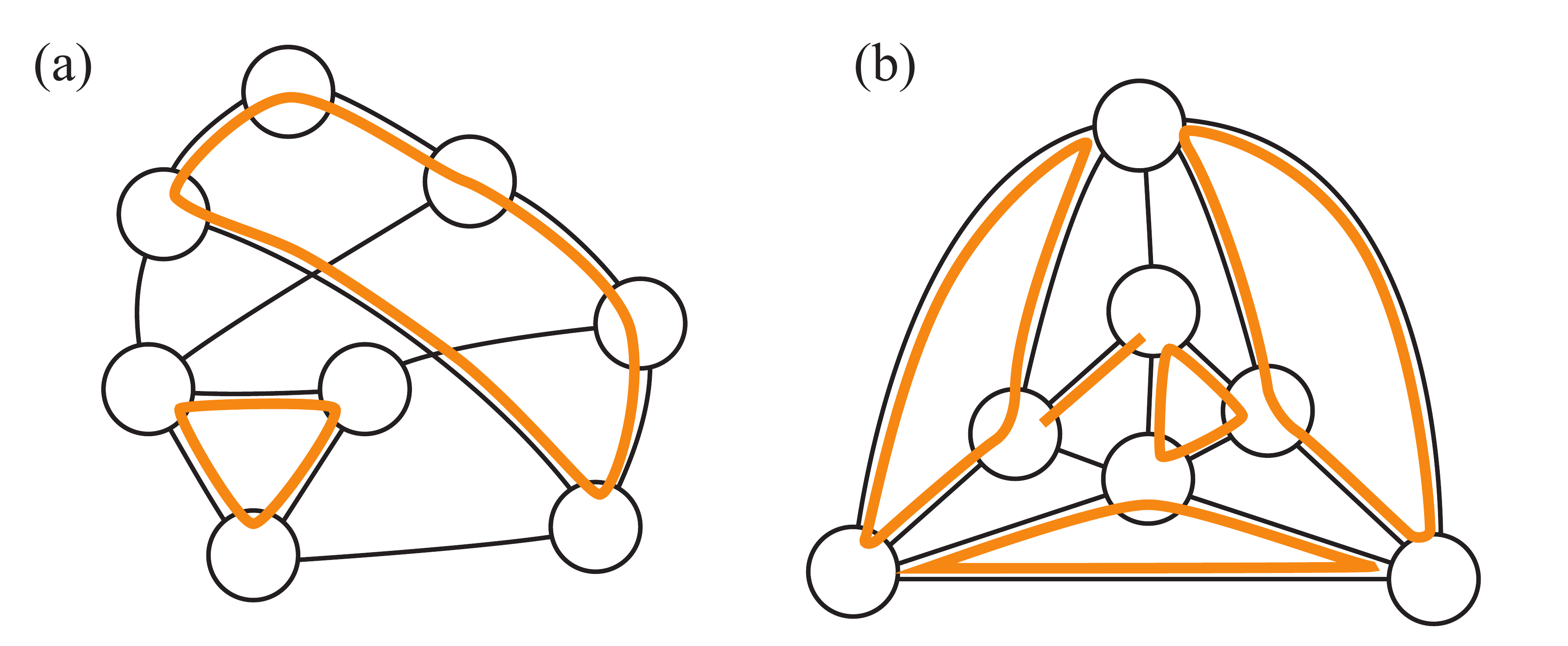}{
(Color online)
Examples of the cycle-chain refutation.
(a) All the vertices in the graph given in \fref{mis-ex} is covered once by a 5-cycle and a 3-cycle, which means $N^*\leq 3$.
(b) All the vertices are covered twice by four 3-cycle and a single chain, which means $N^*\leq 5/2$.
Because $N^*$ is integer, this is equivalent to $N^*\leq 2$.
}{refute}

\para{Cycle-chain refutation}
It should be noted that we should not aim for a hardness proof regarding {\it any} type of refutation, since that will imply P$\neq$NP, 
which is believed to be true but proof is currently out of reach. 
On the other hand, proving hardness for only {\it one particular} algorithm 
may raise questions on the generality of the result.
Thus, to discuss the hardness of computation in as general a way as possible, we specify the {\it method} of refutation, so that any algorithm that employs the same strategy will follow our result.
We here introduce the idea of {\it cycle-chain refutation}, which captures the refutation ability of various algorithms for MIS. 
In the following, we shall first explain the procedure of the cycle-chain refutation, and then argue that it captures the ability of many known algorithms.
In the cycle-chain refutation, vertices of the graph are covered by cycles (closed loop path), chains (pair of neighboring vertices), and single vertices such that any vertex is covered by them the same number of times.
Examples are drawn in \fref{refute}.
In \fref{refute}.(a), all vertices are covered by cycles just once, while in \fref{refute}.(b), all vertices are covered exactly twice.
A cycle with length $2k+1$ or $2k$ can have at most $k$ vertices which belong to a given independent set, and in a chain or a single vertex, there could be at most one.
The cycle-chain refutation draws an upper bound of the independent number by combining a coverage of vertices and the above fact.
From \fref{refute}.(a), any independent set has at most two vertices on the 5-cycle and one vertex on the 3-cycle, which means that the independent number is bounded above by three.
From \fref{refute}.(b), any independent set has at most four vertices on the four 3-cycle and one vertex on the chain. All vertices are covered twice, which means that the independent number is bounded above by 5/2.
Since the independent number is always an integer, we find that the independent number of \fref{refute}.(b) is bounded above by two.
We denote by $\Ncc$ the best upper bound of the independent number obtained by the cycle-chain refutation.
Note that there indeed exist independent sets with size three for \fref{refute}.(a) and with size two for \fref{refute}.(b), and thus the cycle-chain refutation works tightly for these examples.

It should be noted that this cycle-chain refutation is not an algorithm, since we do not specify the way how this refutation is obtained.
While the presence of a cycle-chain refutation does not necessarily imply the existence of an efficient algorithm to find it, the absence of cycle-chain refutation will imply that all algorithms relying on it cannot conduct an efficient refutation.

Linear programming relaxation and kernelization are two common algorithms used for MIS~\cite{NT, Cygbook}.
Importantly, they could both be naturally regarded as a refutation process which upper bounds the independent number of a given instance.
The refutation of both algorithms relies on a property of MIS called {\it half-integrality}, which is only caused by odd-length cycles. 
Since our cycle-chain refutation directly deals with the information of the odd-length cycles, its refutation ability is at least as powerful as linear programming relaxation and kernelization.
In the following, we will consider graphs with all vertices having degree 2 or 3. 
In this case, linear programming relaxation only yields a trivial upper bound of $N/2$ for the independent number,
and thus the cycle-chain refutation always provides a tighter upper bound. 
Furthermore, different algorithms such as the belief propagation and the leaf removal algorithm, 
which have been analyzed from the perspective of statistical mechanics before, 
are known to fail in exactly the same region as the linear programming relaxation fails~\cite{TH14, TH}. 
This means that the cycle-chain refutation is at least as powerful as the refutation ability of various algorithms for MIS.

\para{Construction of the graph sequence}
We now construct a graph sequence of hard instances.
The graph has $p$ vertices labeled as $\{ 0,1,\ldots, p-1\}$ with a prime $p$.
Each vertex $x$ is connected to $x\pm 1$ and $-x^{-1}$ (mod $p$), where we let $0^{-1}=0$ and thus the vertex 0 has a self loop~\cite{note1}.
We call this graph as {\it inverse graph}.
The inverse graph with $p=11$ is depicted in \fref{expander}.
This graph turns out to be a well-known example of an expander graph~\cite{HLW}.
Expander graphs are graph sequences with large expansion ratios~\cite{ex-ratio} and thus random walks on them mix rapidly.

We shall show the following two relations for the sequence of the inverse graphs:
\balign{
\lim_{p\to \infty}\frac{\Ncc}{N}&=\frac12, \lb{main-1} \\
\lim_{p\to \infty}\frac{N^*}{N}&<\frac12. \lb{main-2}
}
The first part \eqref{main-1} shows that the cycle-chain refutation only reproduces a trivial upper bound $1/2$ in the large size limit.
On the other hand, the second part \eqref{main-2} shows that the independent ratio is strictly less than $1/2$.
Combining these two relations, we conclude that any algorithm based on the cycle-chain refutation cannot solve MIS efficiently (i.e., we need a brute-force type search for refutation, which could take exponentially long time in general).

\figin{7cm}{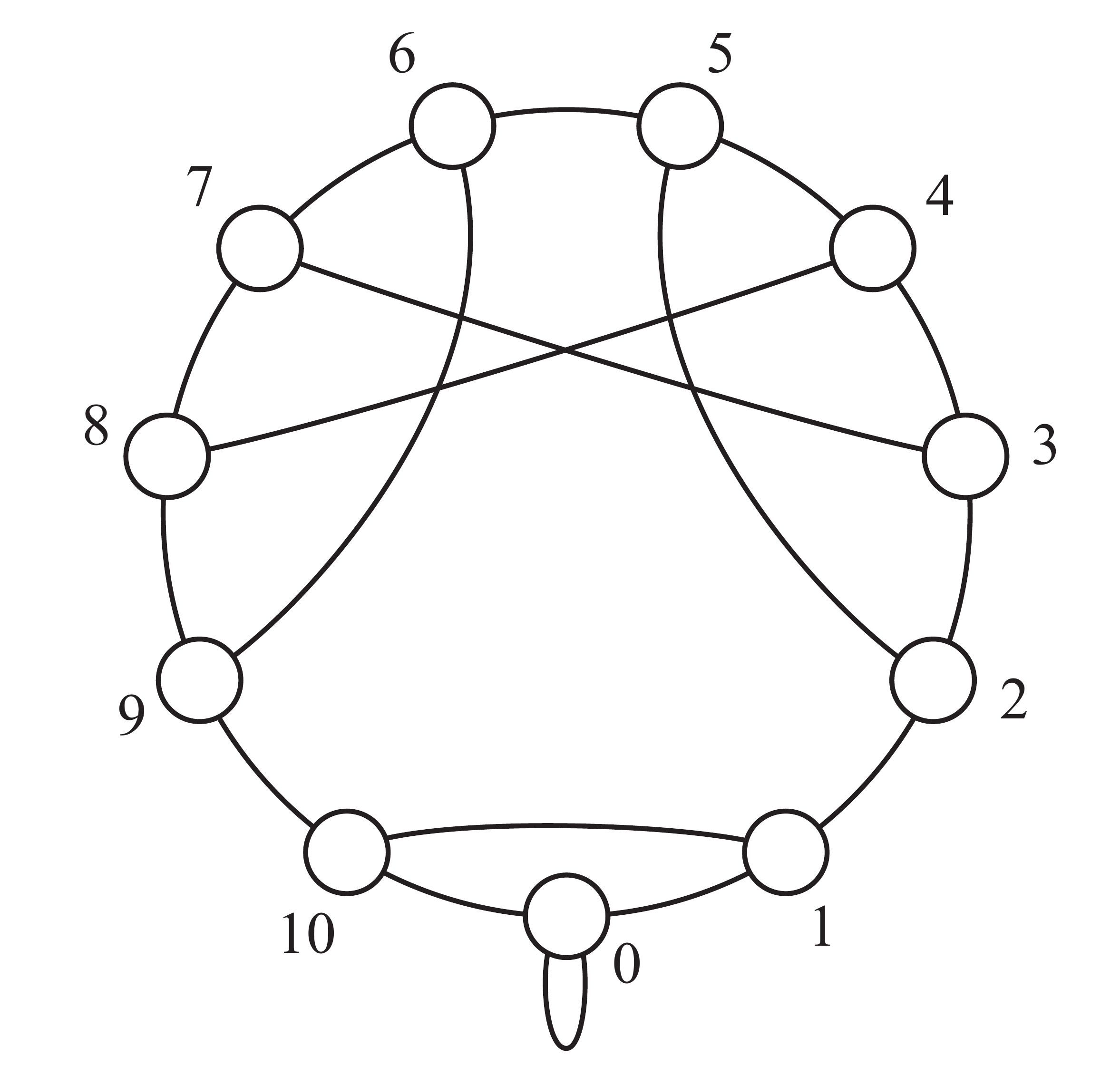}{
An example of the inverse graph with $p=11$.
The vertex $x$ is connected to $x+1$, $x-1$ and $-x^{-1} \mod p$. 
}{expander}

\para{Part 1: Lower bound on refutation}
We first show that for any $\ep>0$ there exists $p'$ such that $\Ncc/N>1/2-\ep$ holds for any $p>p'$.
Consider a coverage of a given cycle-chain refutation.
Let $\no_k$ be the number of odd-length cycles with length $2k+1$, $\ne$ be the number of chains (where we decompose all even-length cycles into chains), $\nv$ be the number of single vertices.
We then have
\eq{
\Ncc = \left\lfloor \frac{\sum_k k\no_k +\ne +\nv}{\sum_k (2k+1)\no_k+2\ne +\nv}\cdot N \right\rfloor,
}
where $\lfloor \cdot \rfloor$ is the floor function.
The denominator $\sum_k (2k+1)\no_k+2\ne +\nv$ represents the cumulative total number of vertices covered by cycles, chains, and single vertices, and the numerator $\sum_k k\no_k +\ne +\nv$ represents the upper bound of the cumulative total number of vertices in an independent set counted by the cycle-chain refutation.
If all the vertices are covered $m$ times, the denominator is equal to $mN$.
This relation confirms the fact that only small odd-length cycles (i.e., small odd $k$) are responsible for good refutations with tighter bounds. 

A crucial property of the inverse graph is that the number of small odd-length cycles in this graph is rigorously bounded.
To explain this, we introduce a symbol sequence to describe a directed path.
A single directed jump from $x$ to $x+1$, $x-1$, $-x^{-1}$ is denoted by $+$, $-$, $R$, respectively, and a directed path from a given initial vertex is written as a sequence of symbols $\{ +,-,R\}$.
For example, a directed path $2\to 3\to 7\to 6\to 5\to 4\to 8$ in \fref{expander} is expressed as the sequence $[+R---R]$ with the initial vertex 2.
A cycle can be regarded as a directed path with the same initial and final state $x$.
By choosing a proper initial state $x$ and a proper direction, any cycle can be described in the form of $[+\cdots R]$ without loss of generality.
We refer to such a sequence as {\it cycle sequence}.

As an example, let us consider a 5-cycle described by $[++R-R]$ (see \fref{short-cycle}).
By construction, the initial state of this cycle $x$ is the solution of
\eqa{
-\frac{1}{-\frac{1}{x+2}-1}\equiv x \mod p.
}{++R-R}
Because the left hand side is $(x+2)/(x+3)$, this equation is quadratic and has at most two solutions.
This means that the inverse graph contains at most two 5-cycles expressed as $[++R-R]$.
For any cycle sequence, the corresponding equation has at most two solutions, and hence the cycle described by this cycle sequence appears at most twice in a given inverse graph.
The exact number of cycles described by a particular cycle sequence in a given inverse graph could be obtained by calculating the quadratic residue~\cite{Cbook}.
(For example, \eref{++R-R} is transformed into a quadratic equation $y^2\equiv 3 \mod p$ with $y=x+2$, which has two solutions for $p\equiv 1, 11 \mod 12$ and no solution for $p\equiv 5,7 \mod 12$.
See also Supplemental Material~\cite{SM})

\figin{7cm}{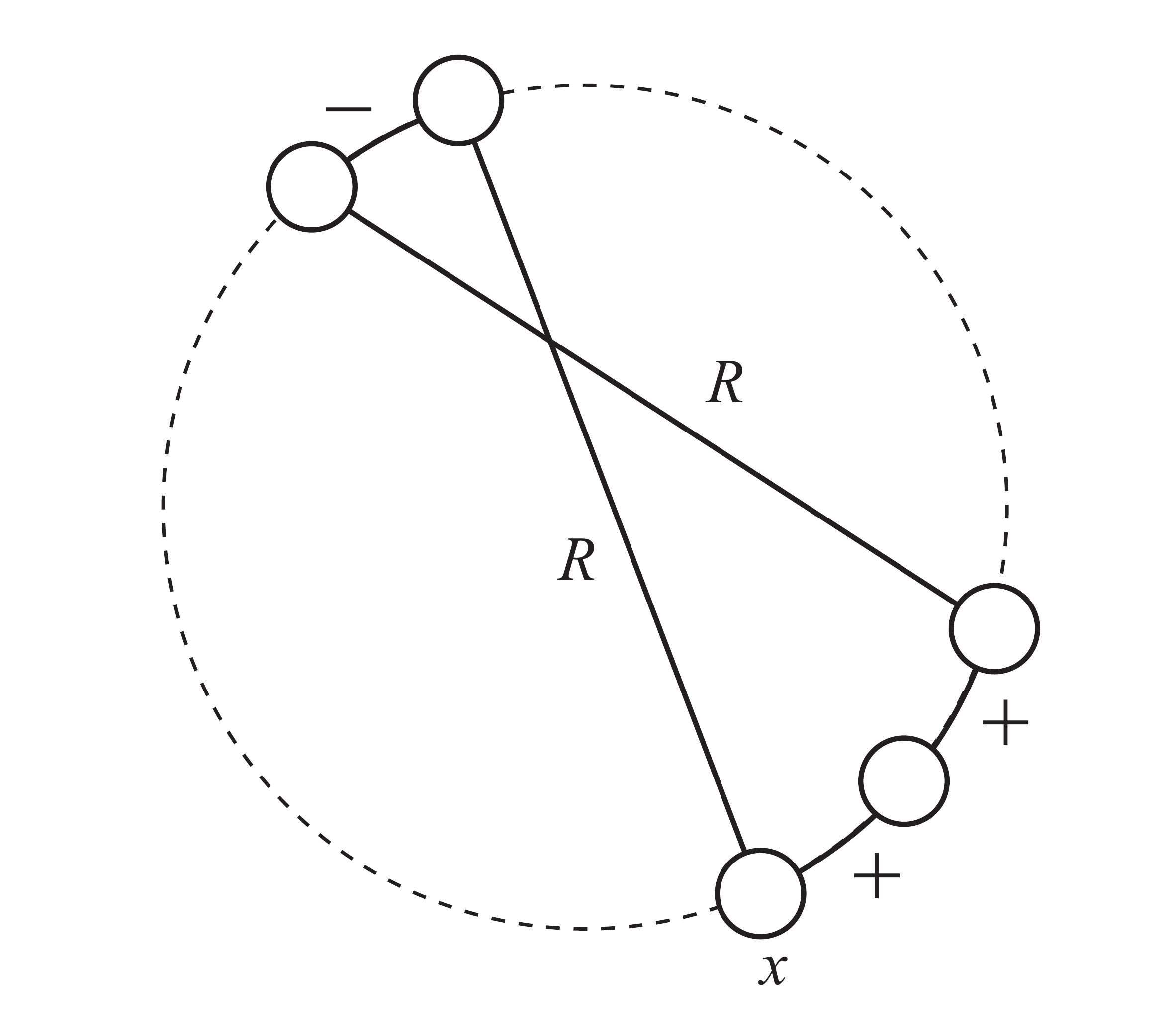}{
Schematic diagram of a possible pass of $[++R-R]$ with the initial vertex $x$.
}{short-cycle}

Because the pairs $+-$, $-+$ and $RR$ do not appear in cycle sequences, the number of cycle sequences with length $2k+1$ is bounded above by $2^{2k-1}$.
This directly implies that $2k+1$ cycles appear in the inverse graph for any $p$ at most $2\cdot 2^{2k-1}=2^{2k}$ times.
Hence, in any cycle-chain refutation, the number of vertices covered by odd-length cycles with length less than $2k'$ is bounded above by a constant independent of $p$: $a(k'):=\sum_{k=1}^{k'-1} (2k+1)2^{2k}$.
In other words, short odd-length cycles cover only a small fraction of vertices for large $p$.
Then, by defining $b(k'):=\sum_{k=1}^{k'-1} k2^{2k}$ and setting as
\balign{
k'\geq&\frac12 \( \frac1\ep -1\) , \\
N=p\geq p':=&a(k')+\frac{a(k')-2b(k')+2}{\ep}
}
for a given $\ep>0$, we have a lower bound of $\Ncc$:
\balign{
\Ncc \geq&\frac{\sum_k k\no_k +\ne +\nv}{\sum_k (2k+1)\no_k+2\ne +\nv}\cdot N -1 \nt \\
\geq &\frac{b(k')+k'\cdot  \frac{N-a(k')}{2k'+1}}{a(k')+(2k'+1) \frac{N-a(k')}{2k'+1}}\cdot N-1
\geq N\( \frac 12 -\ep \) .
}
In the second line, because at most $a(k')$ vertices are covered by odd-length cycles with length less than $2k'$, $\Ncc$ is bounded below by the case that $a(k')$ vertices are covered by short odd-length cycles with length less than $2k'$ and other vertices are covered by cycles with length $2k'+1$.
Because $\ep$ is arbitrary, this relation is equivalent to the desired relation \eqref{main-1}.

\para{Part 2: Upper bound on independent ratio}
We next show two arguments, one analytical and the other numerical, that the true independent ratio is strictly less than half.
First, we exactly compute the independent ratio of the inverse graph up to $p=311$, using a refined brute force search algorithm~\cite{ore}.
We remark that the computation time by this algorithm increases exponentially with the size $p$.
The obtained results are shown in \fref{ind-ratio}, where the independent ratio is plotted against the system size $N=p$. 
The plot suggests that although the independent ratio fluctuates, it converges to a value around 0.46, which is strictly less than half.

\figin{8cm}{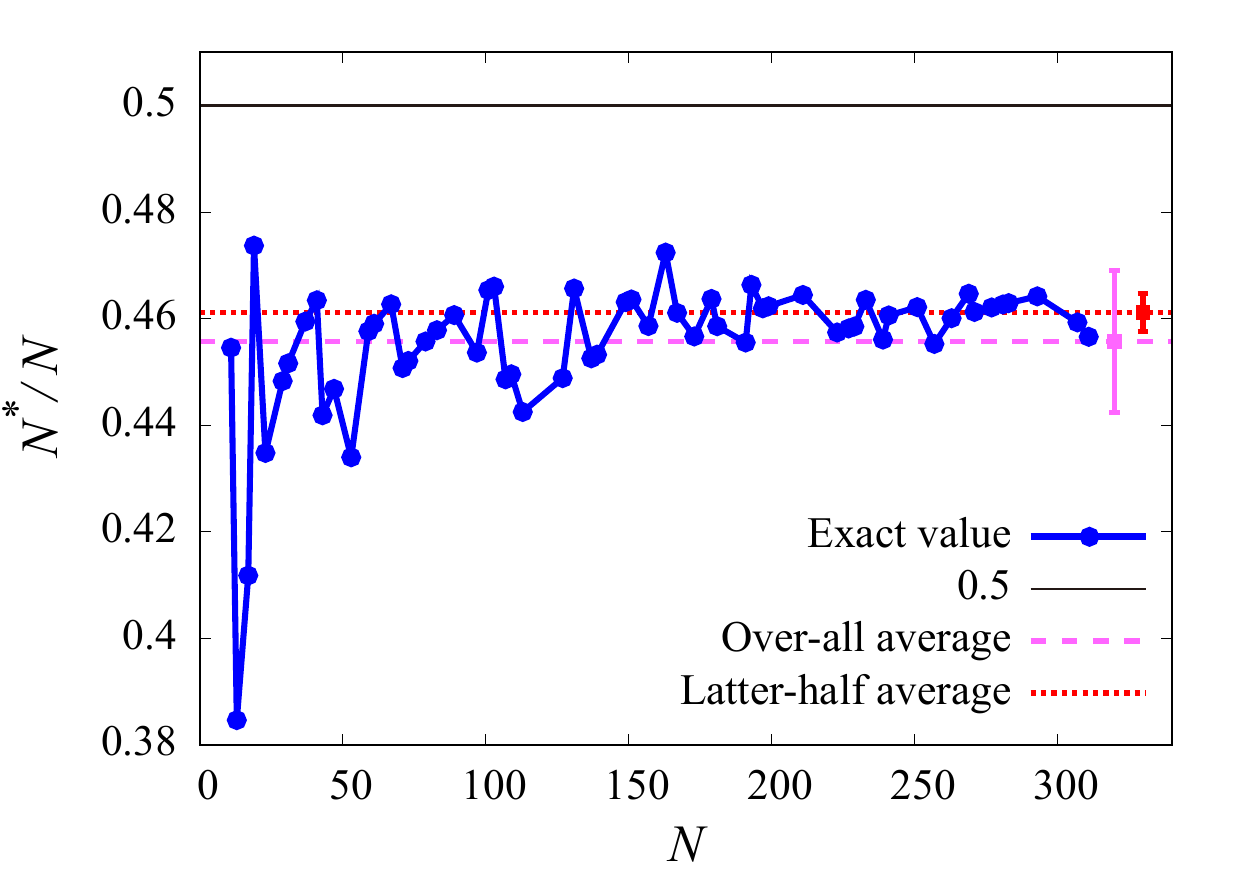}{
(Color online)
Exactly computed independent ratio for the inverse graph from $p=11$ to $p=311$.
Two dotted lines show the average value of the data, either using all of it or only the latter half.
Both of the error bars do not exceed 0.48, and becomes smaller for the latter half, which strongly suggests a convergence to a value strictly less than 0.5. 
}{ind-ratio}

We provide another analytic argument supporting our claim.
Let $A$ be the $N\times N$ normalized adjacency matrix of a $d$-regular graph, where $A_{ij}=1/d$ if there is an edge between the vertices $i$ and $j$ in the graph and $A_{ij}=0$ otherwise.
We set $A_{ii}=0$.
We denote the eigenvalues of $A$ with decreasing order by $\lmd_1\geq \lmd_2\geq \cdots \geq \lmd_N$.
If the graph is $d$-regular, it is known that $\lmd_1=1$, $\lmd_N\geq -1$ and the equality holds if and only if the graph is bipartite~\cite{BHbook}.
In addition, the independent ratio of a $d$-regular graph is bounded above by the smallest eigenvalue, which is called Hoffman's bound~\cite{BHbook}:
\eq{
\frac{N^*}{N}\leq \frac{-\lmd_N}{1-\lmd_N}.
}
Thus, obtaining a good lower bound of $\lmd_N$, we have a good bound of the independent ratio.
It is noteworthy that the second largest eigenvalue in the absolute sense $\lmd :=\max(\lmd_2, -\lmd_N)$ characterizes the speed of mixing through random walks on the graph.
The larger the gap $1-\lmd$ is, the quicker the probability distribution equilibrates.
For example, a review paper on the expander graph~\cite{HLW} shows that the inverse graph satisfies $\lmd<1-10^{-4}$ for large $N$.
This relation suggests that the independent ratio for sufficiently large $N$ is strictly less than half:
\eq{
\frac{N^*}{N}\leq 0.499975<0.5.
}
While a tighter bound of $\lmd_N$ should give us a better bound on the independent ratio, this already suffices for our argument.


\para{Discussion}
We explicitly constructed hard instances (an infinite sequence of graphs) of the maximum independent set problem (MIS).
For this graph, the cycle-chain refutation, which is a plausible method of refutation, only provides the upper bound of the independent ratio as $1/2$, while the true independent ratio is strictly less than $1/2$.
This difference implies that the hardness stems from not local but global structures of the graph.
To our best knowledge, this is the first explicit deterministic construction of a hard instance in optimization problems.

It is noteworthy that the inverse graph shares various properties with $K$-regular random graphs with $K \geq 3$, which are known to be typically hard for MIS~\cite{Zhou}.
For instance, it is known that regular random graphs are expander graphs~\cite{HLW} and have few short cycles~\cite{Wor81} with high probability.
The expected independent ratio of 3-regular random graphs is almost surely larger than $6\ln (2/3)-2\simeq 0.4328$~\cite{FS94, Wor95} and smaller than 0.458~\cite{Dal09}, which is close to the numerical result of the inverse graph, which is also almost 3-regular.
These facts lead us to anticipate that the introduced inverse graph shares the {\it typical} properties of random graphs, which are hard instances of MIS.
Recent studies on computational hardness from the viewpoint of statistical mechanics have paid much attention to the connection between the energy landscape of the solution space and the hardness of problems~\cite{Krz07, ZK07, AC, Zde, ZK10, Bra}, which also resulted in inspiring efficient algorithms~\cite{MRS, MPR, Bal}.
A concrete example of a hard instance will help our investigation of the deep structure of computational hardness and serve as a benchmark for analytical evaluation of algorithms, and would further provide insights on average-case hardness and various phenomena arising from it.

\para{Acknowledgement}
We thank Koji Hukushima, Yoshiyuki Kabashima, Tomoyuki Obuchi, and Satoshi Takabe for fruitful discussions.
NS was supported by Grant-in-Aid for JSPS Fellows JP17J00393. 

\clearpage

\makeatletter
\long\def\@makecaption#1#2{{
\advance\leftskip1cm
\advance\rightskip1cm
\vskip\abovecaptionskip
\sbox\@tempboxa{#1: #2}%
\ifdim \wd\@tempboxa >\hsize
 #1: #2\par
\else
\global \@minipagefalse
\hb@xt@\hsize{\hfil\box\@tempboxa\hfil}%
\fi
\vskip\belowcaptionskip}}
\makeatother
\newcommand{\vo}{\upsilon}
\newcommand{\midskip}{\vspace{3pt}}

\setcounter{equation}{0}
\def\theequation{A.\arabic{equation}}

\begin{widetext}

\begin{center}
{\bf \Large Supplemental Materials for ``\titlename"}

\bigskip
Naoto Shiraishi and Jun Takahashi
\end{center}




\bigskip\noindent
{\bf \large Table of presence or absence of short cycles with odd length}
\midskip

\begin{table}[h]
\begin{tabular}{|c|l|l|l|l|l|}\hline
Length & Sequence & Quadratic equation & two solutions &no solution &others \\ \hline \hline
3 & $[++R]$& & & & $x=-1$ is a unique solution for any $p$ \\ \hline
&$[++++R]$& $y^2\equiv 3 \mod p$& $p\equiv 1,11 \mod 12$ & $p\equiv 5,7 \mod 12$ & \\ \cline{2-6}
5&$[++R-R]$& $y^2\equiv 3 \mod p$& $p\equiv 1,11 \mod 12$ & $p\equiv 5,7 \mod 12$ & \\ \cline{2-6}
&$[++R+R]$& $y^2\equiv -1 \mod p$& $p\equiv 1 \mod 4$ & $p\equiv 3 \mod 4$ & \\ \hline
&$[++++++R]$& $y^2\equiv 2 \mod p$& $p\equiv 1,7 \mod 8$ & $p\equiv 3,5 \mod 8$ & \\ \cline{2-6}
&$[++++R+R]$& &  &  & $x=-2$ is a unique solution for any $p$ \\ \cline{2-6}
&$[++++R-R]$& $y^2\equiv -1 \mod p$& $p\equiv 1 \mod 4$ & $p\equiv 3 \mod 4$ & \\ \cline{2-6}
&$[+++R++R]$& $y^2\equiv 15 \mod p$&  & & \\ \cline{2-6}
7&$[+++R--R]$& $y^2\equiv 3 \mod p$& $p\equiv 1,11 \mod 12$ & $p\equiv 5,7 \mod 12$ & \\ \cline{2-6}
&$[++R+R-R]$& &  &  & no solution for any $p$ \\ \cline{2-6}
&$[++R-R+R]$& &  &  & no solution for any $p$ \\ \cline{2-6}
&$[++R+R+R]$& $y^2\equiv 3 \mod p$& $p\equiv 1,11 \mod 12$ & $p\equiv 5,7 \mod 12$ & \\ \cline{2-6}
&$[++R-R-R]$& $y^2\equiv 3 \mod p$& $p\equiv 1,11 \mod 12$ & $p\equiv 5,7 \mod 12$ & \\ \hline
&$[++++++++R]$& $y^2\equiv 15 \mod p$&  & & \\ \cline{2-6}
&$[++++++R+R]$& $y^2\equiv 3 \mod p$& $p\equiv 1,11 \mod 12$ & $p\equiv 5,7 \mod 12$ & \\ \cline{2-6}
&$[++++++R-R]$& $y^2\equiv 15 \mod p$&  & & \\ \cline{2-6}
&$[+++++R++R]$& $y^2\equiv 35 \mod p$&  & & \\ \cline{2-6}
&$[+++++R--R]$& $y^2\equiv 15 \mod p$&  & & \\ \cline{2-6}
&$[++++R+++R]$& $y^2\equiv 3 \mod p$& $p\equiv 1,11 \mod 12$ & $p\equiv 5,7 \mod 12$ & \\ \cline{2-6}
&$[++++R---R]$& $y^2\equiv 6 \mod p$&  & & \\ \cline{2-6}
&$[++++R+R+R]$& &  &  & no solution for any $p$ \\ \cline{2-6}
&$[++++R-R-R]$& &  &  & no solution for any $p$ \\ \cline{2-6}
&$[++++R+R-R]$& $y^2\equiv 15 \mod p$&  & & \\ \cline{2-6}
&$[++++R-R+R]$& $y^2\equiv 15 \mod p$&  & & \\ \cline{2-6}
&$[+++R++R+R]$& $y^2\equiv -1 \mod p$& $p\equiv 1 \mod 4$ & $p\equiv 3 \mod 4$ & \\ \cline{2-6}
&$[+++R++R-R]$& $y^2\equiv 6 \mod p$&  & & \\ \cline{2-6}
&$[+++R--R+R]$& $y^2\equiv 15 \mod p$&  & & \\ \cline{2-6}
9&$[+++R--R-R]$& $y^2\equiv 2 \mod p$& $p\equiv 1,7 \mod 8$ & $p\equiv 3,5 \mod 8$ & \\ \cline{2-6}
&$[+++R+R++R]$& $y^2\equiv -1 \mod p$& $p\equiv 1 \mod 4$ & $p\equiv 3 \mod 4$ & \\ \cline{2-6}
&$[+++R+R--R]$& $y^2\equiv 15 \mod p$&  & & \\ \cline{2-6}
&$[+++R-R++R]$& $y^2\equiv 6 \mod p$&  & & \\ \cline{2-6}
&$[+++R-R--R]$& $y^2\equiv 2 \mod p$& $p\equiv 1,7 \mod 8$ & $p\equiv 3,5 \mod 8$ & \\ \cline{2-6}
&$[++R++R++R]$& &  &  & $x=-1$ is a unique solution for any $p$ \\ \cline{2-6}
&$[++R++R--R]$& $y^2\equiv 6 \mod p$&  & & \\ \cline{2-6}
&$[++R+R+R+R]$& &  &  & no solution for any $p$ \\ \cline{2-6}
&$[++R+R-R+R]$& &  &  & no solution for any $p$ \\ \cline{2-6}
&$[++R+R+R-R]$& $y^2\equiv -1 \mod p$& $p\equiv 1 \mod 4$ & $p\equiv 3 \mod 4$ & \\ \cline{2-6}
&$[++R+R-R-R]$& $y^2\equiv 3 \mod p$& $p\equiv 1,11 \mod 12$ & $p\equiv 5,7 \mod 12$ & \\ \cline{2-6}
&$[++R-R+R+R]$& $y^2\equiv -1 \mod p$& $p\equiv 1 \mod 4$ & $p\equiv 3 \mod 4$ & \\ \cline{2-6}
&$[++R-R+R-R]$& $y^2\equiv 6 \mod p$&  & & \\ \cline{2-6}
&$[++R-R-R+R]$& $y^2\equiv 3 \mod p$& $p\equiv 1,11 \mod 12$ & $p\equiv 5,7 \mod 12$ & \\ \cline{2-6}
&$[++R-R-R-R]$& &  &  & $x=-1$ is a unique solution for any $p$ \\ \hline
\end{tabular}
\end{table}

The table in the previous page shows which type of cycle appears in particular inverse graphs.
The table is read as follows.
For example, the number of the 7-cycle described by the cycle sequence $[++++++R]$ is equal to the number of solutions of the equation $y^2\equiv 2 \mod p$, 
which is two If $p\equiv 1,7 \mod 8$ and is none if $p\equiv 3,5 \mod 8$.
In contrast, the 7-cycle described by the cycle sequence $[++++R+R]$ always exists uniquely, and the initial vertex of this sequence is always $x=-2$.
Here, the equation corresponding to this sequence as \eref{++R-R} is not a quadratic equation but a simple first-degree polynomial equation.
In some cases, this equation has no solution for any $p$, which occurs, for example, for the 7-cycle described by the cycle sequence $[++R+R-R]$.

In this table, we leave some columns for quadric residues of composite numbers blank, because the conditions for the solutions to have two/no solutions are relatively complicated.
The explicit conditions for those cases could be obtained in the following way.

\

We here briefly explain a mathematical background behind this calculation.
We first introduce the Legendre's symbol for a prime $p$ and an integer $a$ which are relatively prime:
\eq{
\( \frac ap \) :=
\bcases{
1 &: y^2\equiv a \ (\text{mod } p) \text{ has two solutions} \\
-1 &: y^2\equiv a \ (\text{mod } p) \text{ has no solution}.
}
}
The Legendre's symbol satisfies the following relation
\eq{
\( \frac{ab}{p}\) =\( \frac ap \) \( \frac bp \) ,
}
which could be used to calculate the conditions of composite quadratic residues having two/no solutions.
This relation says that $y^2\equiv ab \mod p$ has two solutions if and only if (i) both $y^2\equiv a \mod p$ and $y^2\equiv b \mod p$ have two solutions, 
or (ii) both $y^2\equiv a \mod p$ and $y^2\equiv b \mod p$ have no solution.
For example, $y^2\equiv 15 \mod p$ has two solutions if and only if (i) both $p\equiv 1,11 \mod 12$ and $p\equiv 1,4 \mod 5$ hold, 
or (ii) both $p\equiv 5,7 \mod 12$ and $p\equiv 2,3 \mod 5$ hold.

To calculate the quadratic residue of primes, we employ the quadratic reciprocity
\eq{
\( \frac pq\) \( \frac qp \) =(-1)^{\frac{p-1}{2}\frac{q-1}{2}},
}
and the following two supplementary laws
\balign{
\( \frac{-1}{p}\) &= (-1)^{\frac{p-1}{2}}, \\
\( \frac{2}{p}\) &= (-1)^{\frac{p^2-1}{8}}.
}
We list the conditions for quadratic residues of small primes and the special case -1 below.

\begin{table}[h]
\begin{tabular}{|l|l|l|}\hline
Quadratic equation & two solutions &no solution  \\ \hline \hline
$y^2\equiv -1 \mod p$& $p\equiv 1 \mod 4$ & $p\equiv 3 \mod 4$  \\ \hline
$y^2\equiv 2 \mod p$& $p\equiv 1,7 \mod 8$ & $p\equiv 3,5 \mod 8$ \\ \hline
$y^2\equiv 3 \mod p$& $p\equiv 1,11 \mod 12$ & $p\equiv 5,7 \mod 12$ \\ \hline
$y^2\equiv 5 \mod p$& $p\equiv 1,4 \mod 5$ & $p\equiv 2,3 \mod 5$ \\ \hline
$y^2\equiv 7 \mod p$& $p\equiv 1,3,9,19,25,27 \mod 28$ & $p\equiv 5,11,13,15,17,23 \mod 28$ \\ \hline
\end{tabular}
\end{table}

\clearpage

\end{widetext}

\end{document}